\documentclass[conference]{IEEEtran}
\usepackage[pdftex]{graphicx}

\usepackage{color}

\newcommand{\regret}{\mathrm{regret}}
\newcommand{\NML}{\mathrm{NML}}

\usepackage{amssymb}
\usepackage{amsmath}
\usepackage{euscript}

\title{Bayesian Properties of Normalized Maximum Likelihood and its Fast Computation}

\author{\IEEEauthorblockN{Andrew Barron}
\IEEEauthorblockA{Department of Statistics\\
Yale University\\
Email: andrew.barron@yale.edu}
\and
\IEEEauthorblockN{Teemu Roos}
\IEEEauthorblockA{Department of Computer Science\\
University Helsinki\\
Email: teemu.roos@cs.helsinki.fi}
\and
\IEEEauthorblockN{Kazuho Watanabe}
\IEEEauthorblockA{Graduate School of Information Science\\
Nara Institute of Science and Technology\\
Email: wkazuho@is.naist.jp}}

\begin{document}
\maketitle
\begin{abstract}
The normalized maximized likelihood (NML) provides the minimax regret
solution in universal data compression, gambling, and prediction, and
it plays an essential role in the minimum description length (MDL)
method of statistical modeling and estimation.  Here we show that the
normalized maximum likelihood has a Bayes-like representation as a
mixture of the component models, even in finite samples, though the
weights of linear combination may be both positive and negative.  This
representation addresses in part the relationship between MDL and
Bayes modeling.  This representation has the advantage of speeding the
calculation of marginals and conditionals required for coding and
prediction applications.
\end{abstract}

\section{Introduction}
For a family of probability mass (or probability density) functions $p(x;\theta)$, also denoted $p_\theta(x)$ or $p(x|\theta)$, for data $x$ in a data space $\cal X$ and a parameter $\theta$ in a parameter set $\Theta$, there is a distinquished role in information theory and statistics for the maximum likelihood measure with mass (or density) function proportional to $p(x;\hat \theta(x))$ obtained from the maximum likelihood estimator $\hat \theta(x)$ achieving the maximum likelihood value $m(x)=\max_\theta \, p(x;\theta)$.  Let $C = \sum_x m(x)$, where the sum is replaced by an integral in the density case.  For statistical models in which $\sum_x m(x)$ is finite (i.e. the maximum likelihood measure is normalizable), this maximum value $m(x)$ characterizes exact solution in an arbitrary sequence (non-stochastic) setting to certain modeling tasks ranging from universal data compression, to arbitrage-free gambling, to predictive distributions with minimax regret.

Common to these modeling tasks is the problem of providing a single non-negative distribution $q(x)$ with $\sum_x q(x) = 1$ with a certain minimax property.  For instance, for the compression of data $x$ with codelength $\log 1/q(x)$, the codelength is to be compared to the best codelength with hindsight $\min_\theta \log 1/p_\theta(x)$ among the codes parameterized by the family.  This ideal codelength is not exactly attainable, because the maximized likelihood $m(x)$ will (except in a trivial case) have a sum than is greater than $1$, so that the Kraft inequality required for unique decodability would not be satisfied by plugging in the MLE.  We must work with a $q(x)$ with sum not greater than $1$. The difference between these actual and ideal codelengths is the (pointwise)  regret
$$\regret(q,x) = \log \frac{1}{q(x)} - \min_\theta \log \frac{1}{p_\theta (x)}$$
which of course is the same as 
$\regret(q,x) = \log \frac{m(x)}{q(x)}.$
The minimax regret problem is to solve for the distribution $q^*(x)$ achieving 
$$\min_q \max_{x \in \cal X} \regret(q,x)$$
where the minimum is taken over all probability mass functions.  
Beginning with Shtarkov \cite{Shtarkov}, who formulated the minimax regret problem for universal data compression, it has been shown that the solution is given by the normalized maximized likelihood $q^*(x)=\NML(x)$ given by
$$\NML(x) = \frac{m(x)}{C}$$
where $C=C_{Shtarkov}$ is the normalizer given by $\sum_x m(x)$.  This $q^*(x)$ is an equalizer rule (achieving constant regret), showing that the minimax reget is $\log C$.

For settings in which $C=\sum_x m(x)$ is infinite, the maximized likelihood measure $m(x)$ is not normalizable and the minimax regret defined above is infinite.  Nevertheless, one can identify problems of this type in which the maximized likelihood value continues to have a distinguished role.  In particular, suppose the data comes in two parts $(x,x'),x \in {\cal X}, x' \in {\cal X}'$, thought of as initial and subsequent data strings.  Then the maximum likelihood value $m(x,x')=\max_\theta p(x,x';\theta)$ often has a finite marginal $m_{init}(x)= \sum_{x'} m(x,x')$ leading to the conditional NML distribution
$$\mathrm{condNML}(x'|x) = \frac{m(x,x')}{m_{init}(x)}$$
which is non-negative and sums to $1$ over $x' \in {\cal X}'$ for each such conditioning event $x \in {\cal X}$.

Bayes mixtures are used in approximation, and as we shall see, in exact representation of the maximized likelihood measure.  There are reasons for this use of Bayes mixtures when studying properties of logarithmic regret in general and when studying the normalized maximum likelihood in particular.  There are two traditional reasons that have an approximate nature. One is a relationship between minimax pointwise regret and minimax expected regret for which Bayes procedures are known to play a distinquished role. The other is the established role of such mixtures in the asymptotic charaterization of minimax pointwise regret. 

Here we offer up two more reasons for consideration of Bayes mixtures which are based on exact representation of the normalized maximum likelihood.  One is that representation by mixtures provides computational simplification of coding and prediction by NML or conditional NML.   The other is that the exact representation of NML allows determination of which parametric families allow Bayes interpretation with positive weights and which require a combination of positive and negative weights. 

Before turning attention to exact representation of the NML, let's first recall the information-theoretic role in which Bayes mixtures 
arise.
The expected regret (redundancy) in data compression is  $E_{X|\theta} [\log 1/q(X) - \log 1/p(X|\theta)]$, which is a function of $\theta$ giving the expectation of the difference between the codelength based on $q$ and the optimal codelength (the expected difference being a Kullback divergence).  There is a similar formulation of expected regret for the description of $X'$ given $X$ which is the risk of the statistical decision problem with loss specified by Kullback divergence. 

For these two decision problems the procedures minimizing the average
risk are the Bayes mixture distributions and Bayes predictive
distributions, respectively.
In general
admissibility theory for convex losses like Kullback divergence, the
only procedures not improvable in their risk functions are Bayes and
certain limits of Bayes procedures with positive priors.
For minimax expected redundancy, with min over $q$ and max over $\theta$, the minimax solution is characterized using the maximin average redundancy, which calls for a least favorable (capacity achieving) prior \cite{Haussler}.

The maximum pointwise regret $\max_x \log (m(x)/q(x))$ provides an upper bound on $\max_\theta E_{X|\theta} \log m(X)/q(X)$ as well as an upper bound on the maximum expected redundancy.  It is for the max over $\theta$ problems that the minimax solution takes the form of a Bayes mixture.  So it is a surprise that the max over $x$ form also has a mixture representation as we shall see.

The other traditional role for Bayes mixtures in the study of the NML arises in asymptotics \cite{wrm}. Suppose $x=(x_1,x_2,\ldots, x_n)$ is a string of outcomes from a given alphabet. Large sample approximations for smooth families of distributions show a role for sequences of prior distributions $W_n$  with densities close to Jeffreys prior $w(\theta)$ taken to be proportional to $|I(\theta)|^{1/2}$ where $I(\theta)$ is the Fisher information.  Bayes mixtures of this type are asymptotically minimax for the expected regret \cite{ClarkeBarron,XieBarron97}, and in certain exponential families Bayes mixtures are simultaneously asymptotically minimax for pointwise regret and expected regret \cite{XieBarron00,TakeuchiBarron98}.  However, in non-exponential families,  it is problematic for Bayes mixtures to be asymptotically minimax for pointwise regret, because there are data sequences for which the empirical Fisher information (arising in the large sample Laplace approximation) does not match the Fisher information, so that Jeffreys prior fails.  The work of \cite{TakeuchiBarron98,TakeuchiBarron13} overcomes this problem in the asymptotic setting by putting a Bayes mixture on a slight enlargement of the family to compensate for this difficulty.  The present work motivates consideration of signed mixtures in the original family rather than enlarging the family.

We turn now to the main finite sample reasons for exploration of Bayes representation of NML.  The first is the matter of computational simplification of the representation of NML by mixtures with possibly signed weights of combination. For any coding distribution $q(x_1,x_2,\ldots,x_n)$ in general, and NML in particular, coding implementation (for which arithmetic coding is the main tool) requires computation of the sequence of conditional distributions of $x_i | x_1,\ldots,x_{i-1}$ defined by the ratios of consecutive marginals for $(x_1,...,x_i)$ for $1\le i \le n$.  This appears to be a very difficult task for normalized maximum likelihood, for which direct methods require sums of size up to $|{\cal X}|^{n-1}$.  Fast methods for NML coding have been developed in specialed settings~\cite{KontkanenMyllymaki}, yet remain intractable for most models.

In contrast, for computation of the corresponding ingredients of Bayes mixtures $mix(x^n)= \int p(x^n | \theta) W(d \theta)$, one can make use of simplifying conditional rules for $p(x_i|x_1,\ldots,x_{i-1},\theta)$, e.g. it is equal to $p(x_i|\theta)$ in the conditionally iid case or $p(x_i|x_{i-1},\theta)$ in the first order Markov case, which multiply in providing $p(x^i |\theta)$ for each $i < n$.  So to compute $mix(x_i |x_1,\ldots,x_{i-1})$ one has ready access to the computation of the ratios of consecutive marginals $mix(x_1,\ldots,x_{i})=\int p(x^i |\theta) W(d  \theta)$, contingent on the ability to do the sums or integrals required by the measure $W$.  Equivalently, one has representation of the required predictive distributions $mix(x_i|x_1,\dots,x_{i-1})$ as a posterior average, e.g. in the iid case it is, $\int p(x_i|\theta) W(d\theta|x_1,\ldots,x_{i-1})$.  So Bayes mixtures permit simplified marginalization (and conditioning) compared to direct marginalization of the NML.

The purpose of the present paper is to explore in the finite sample setting, the question of whether we can take advantage of the Bayes mixture to provide exact representation of the maximized likelihood measures.  That is, the question explored is whether there is a prior measure $W$ such that exactly
$$\max_\theta p(x;\theta) = \int p(x;\theta) W(d \theta).$$
Likewise for strings $x^n=(x_1,\ldots,x_n)$ we want the representation $\max_\theta p(x^n;\theta)= \int p(x^n:\theta)W_n(d\theta)$ for some prior measure $W_n$ that may depend on $n$.  Then to perform the marginalization required for sequential prediction and coding by maximized likelihood measures, we can get them computationaly easily as $\int p(x^i;\theta) W_n(d\theta)$ for $i\le n$.  We point out that this computational simplicity holds just as well if $W_n$ is a signed (not necessarily non-negative) measure.


In Section II, we give a result on exact representation in the case that the family has a finitely supported sufficient statistic.  In Section III we demonstrate numerical solution in the Bernoulli trials case, using linear algebra solutions for $W_n$ as well as a Renyi divergence optimization, close to optimization of the maximum ratio of the NML and the mixture.


Finally, we emphasize that at no point are we trying to report a negative probability as a direct model of an observable variable.  Negative weights of combination of unobservable parameters are instead arising in representation and calculation ingredients of non-negative probability mass functions of observable quantities. The marginal and predictive distributions of observable outcomes all remain non-negative as they must.

\section{Signed Prior representation of NML}

We say that a function $mix(x)=\int p(x|\theta)W(d\theta)$ is a signed Bayes mixture when $W$ is allowed to be a signed measure, with positive and negative parts, $W_+$ and $W_-$, respectively.  These signed Bayes mixtures may play a role in the representation of $\NML(x)$.
For now, let's note that for strings $x=(x_1,x_2,\ldots,x_n)$ a signed Bayes mixture has some of the same marginalization properties as a proper Bayes mixture.
The marginal for $ (x_1,\ldots,x_i)$ is defined by summing out $x_{i+1}$ through $x_n$.  For the components $p(x^n|\theta)$, the marginals are denoted $p(x_1,\ldots,x_i|\theta)$. These may be conveniently simple to evaluate for some choices of the component family, e.g. i.i.d. or Markov.  Then the signed Bayes mixture has marginals $mix(x_1,\ldots,x_i) = \int p(x_1,\ldots,x_i|\theta) W(d\theta)$.
[Here it is being assumed that, at least for indicies $i$ past some initial value, the $mix_+(x^i)=\int p(x^i|\theta)W_+(d\theta)$ and $mix_-(x^i)=\int p(x^i|\theta)W_-(d\theta)$ are finite, so that the exchange of the order of the integral and the sum producing this marginal is valid.]   Our emphasis will be on cases in which the mixture $mix(x^n)$ is non-negative (that is $mix_-(x^n) \le mix_+(x^n)$ for all $x^n$) and then the marginals will be non-negative as well. Accordingly one has predictive distributions $mix(x_i|x_1,\dots,x_{i-1})$ defined as ratios of consecutive marginals, as long as the conditioning string has $mix(x_1,\ldots,x_{i-1})$ finite and non-zero. It is seen then that  $mix(x_i|x_1,\dots,x_{i-1})$ is a non-negative distribution which sums to $1$, summing over $x_i$ in $\cal X$, for each such conditioning string.  Moreover, one may formally define a possibly-signed posterior distribution such that the predictive distribution is still a posterior average, e.g. in the iid case one still has the representation $\int p(x_i|\theta)W(d\theta|x_1,\ldots,x_{i-1})$.  

We mention that for most families the maximized likelihood
$m^n(x^n)=\max_\theta p(x^n|\theta)$ has a horizon dependence
property, such that the marginals $m_i^n (x_1,\ldots,x_i)$ defined as
$\sum_{x_{i+1},\ldots,x_n} m^n (x_1,\ldots,x_i,x_{i+1},\ldots,x_n)$
remain slightly dependent on $n$ for each $ i \le n$.  In suitable
Bayes approximations and exact representations, this horizon
dependence is reflected in a (possibly-signed) prior $W_n$ depending
on $n$, such that its marginals $m_i^n (x_1,\ldots,x_i)$ take the form
$\int p(x_1,\ldots,x_i|\theta) W_n(d\theta)$. (Un)achievability of
asymptotic minimax regret without dependency on the horizon was
characterized as a conjecture in~\cite{wrm}.  Three models within
one-dimensional exponential families are exceptions to this horizon
dependence~\cite{bghhk}.  It is also of interest that, as shown in
\cite{bghhk}, those horizon independent maximized likelihood measures
have exact representation using horizon independent positive priors.

Any finite-valued statistic $T=T(x^n)$ has a distribution
$p_T(t|\theta)=\sum_{x^n:T(x^n)=t} p(x^n|\theta)$.  It is a sufficient
statistic if there is a function $g(x^n)$ not depending on $\theta$
such that the likelihood factorizes as
$p(x^n|\theta)=g(x^n)p_T(T(x^n)|\theta)$.  
If the statistic $T$ takes
on $M$ values, 
we may 
regard the distribution of $T$ as a
vector in the positive orthant of $R^M$, with sum of coordinates 
equal to $1$.  For example if $X_1,\ldots, X_n$ are
Bernoulli($\theta$) trials then $T=\sum_{i=1}^n X_i$ is well known to
be a sufficient statistic having a Binomial($n,\theta$) distribution
with $M=n+1$.

The main point of the present paper is to explore the ramifications of the following simple result.

{\bf {Theorem}:}  {\emph {Signed-Bayes representation of maximized likelihood}.} Suppose the parametric family $p(x^n|\theta)$ , with $x^n$ in ${\cal X}^n$ and $\theta$ in $\Theta$, has a sufficient statistic $T(x^n)$ with values in a set of cardinality $M$, where $M=M_n$ may depend on $n$.  Then for any subset $\Theta_M =\{\theta_1,\theta_2,\ldots,\theta_M\}$ for which the distributions of $T$ are linearly independent in $R^M$, there is a possibly-signed measure $W_n$ supported on $\Theta_M$, with values $W_{1,n},W_{2,n},\ldots,W_{M,n}$, such that $m(x^n)=\max_{\theta} p(x^n|\theta)$ has the representation
$$m(x^n) = \int p(x^n|\theta)W_n(d\theta)= \sum_{k=1}^M p(x^n|\theta_k) W_{k,n}.$$

{\bf {Proof}:}  By sufficiency, it is enough to represent $m_T(t)=\max_\theta p_T(t|\theta)$ as a linear combination of $p_T(t|\theta_1),p_T(t|\theta_2),\ldots,p_T(t|\theta_M)$, which is possible since these are linearly independent and hence span $R^M$. 

\vspace{.1cm}

{\bf {Remark 1}:}  Consequently, the task of computation of the marginals (and hence conditionals) of maximized likelihood needed for minimax pointwise redundancy codes is reduced from the seemingly hard task of summing over ${\cal X}^{n-i}$ to the much simpler task of computing the sum of $M$ terms 
$$m_i^n(x_1,\ldots,x_i) = \sum_{k=1}^M p(x_1,\ldots,x_i|\theta_k) W_{k,n}.$$

{\bf {Remark 2}:} Often the likelihood ratio $p(x^n|\theta)/p(x^n|\hat
\theta)$ simplifies, where $\hat \theta$ is the maximum likelihood
estimate.  For instance, in i.i.d. exponential families it takes the
form $\exp\{-nD(\hat\theta||\theta)\}$ where $D(\theta||\theta')$ is
the relative entropy between the distributions at $\theta$ and at
$\theta'$.  So then, dividing through by $p(x^n|\hat \theta)$, the
representation task is to find a possibly-signed measure $W$ such that
the integral of these $e^{-nD}$ is constant for all possible values of
$\hat \theta$, that is,
$\int e^{-nD(\hat\theta||\theta)} W_n(d\theta) = 1.$
The $\hat \theta$ will only depend on the sufficient statistic so this is a simplified form of the representation.

{\bf {Remark 3}:} Summing out $x^n$ one sees that the Shtarkov value $C_{Shtarkov} = \sum_{x^n} m(x^n)$ has the representation $C_{Shtarkov} = \sum_{k=1}^M W_{k,n}$.  That is, the Shtarkov value matches the total signed measure of
$\Theta$.  When $C_{Shtarkov}$ is finite, one may alternatively divide out $C_{Shtarkov}$ and provide a representation
$\NML(x^n) = \int p(x^n|\theta) W_n(d\theta)$
in which the possibly-signed prior has total measure $1$.

{\bf {Remark 4}:} In the Bernoulli trials case, the likelihoods are proportional to $[\theta/(1-\theta)]^T (1-\theta)^n$.  The $(1-\theta)^n$ can be associated with the weights of combination.  To see the linear independence required in the theorem, it is enough to note that the vectors of exponentials $(e^{\eta t}: t=1,2,\ldots,n)$ are linearly independent for any $n$ distinct values of the log odds $\eta =\log[\theta/(1-\theta)]$. The roles of $\theta$ and $1-\theta$ can be exchanged in the maximized likelihood, and, correspondingly, the representation can be arranged with a prior symmetric around $\theta=1/2$.  Numerical selections are studied below.

\section{Numerical Results}
\label{sec:numerics}

We now proceed to demonstrate the discussed mixture representations.

\subsection{A trivial example where negative weigths are required}

We start with a simple illustration of a case where negative weights
are required. Consider a single observation of a ternary random
variable $X\in\{1,2,3\}$ under a model consisting of three 
probability mass functions, $p_1,p_2,p_3$, defined as follows.
$$
p_1 = \left({1\over 2},{1\over 2},0\right), \quad
p_2 = \left(0,{1\over 2},{1\over 2}\right), \quad
p_3 = \left({2\over 7}, {3\over 7}, {2\over 7}\right).
$$ 
The maximum likelihood values are given by $m(x) = \max_\theta
p(x;\theta) = 1/2$ for all $x$ since for all $x\in\{1,2,3\}$ the
maximum, $1/2$, is achieved by either $p_1$ or $p_2$ (or both). The
NML distribution is therefore the uniform distribution $\NML(x) =
(1/3, 1/3, 1/3)$.

An elementary solution for the weights $W_1, W_2, W_3$ such that
$mix(x) = \NML(x)$ for all $x$ yields $W = (-2/3,-2/3,7/3)$. The
solution is unique implying that in particular there is no weight
vector that achieves the matching with only positive weights. 

\subsection{Bernoulli trials: Linear equations for $W_n$ with fixed $\theta$}

For $n$ Bernoulli trials, the sufficient statistic $T=\sum_{i=1}^n
X_i$ takes on $M=n+1$ possible values. We discuss two alternative
methods for finding the prior. First, as a direct application of
linear algebra, we choose a set of $M$ fixed parameter values
$\theta_1,\ldots,\theta_M$ and obtain the weights by solving a system
of $M$ linear equations. By Remark 4 above, any combination of
distinct $\theta$ values yields linearly independent distributions of
$T$ and a signed-Bayes representation is guaranteed to exist. However,
the choice of $\theta$ has a strong effect on the resulting weights
$W_{k,n}$.  

We consider two alternative choices of $\theta_k, k \leq n+1$: first,
a uniform grid with $\theta_k = (k-1)/n$, and second, a grid with
points at $\theta_k = \sin^2((k-1)\pi / 2n)$.  The latter are the quantiles of 
the Beta(1/2,1/2) distribution (also known as the arcsine law), which is the 
Jeffreys prior motivated by the asymptotics discussed in the introduction.  

\begin{figure}
\footnotesize\sc
\hspace*{17mm}Uniform grid\hspace*{28mm}$\sin^2$ grid\\
  \includegraphics[width=\columnwidth]{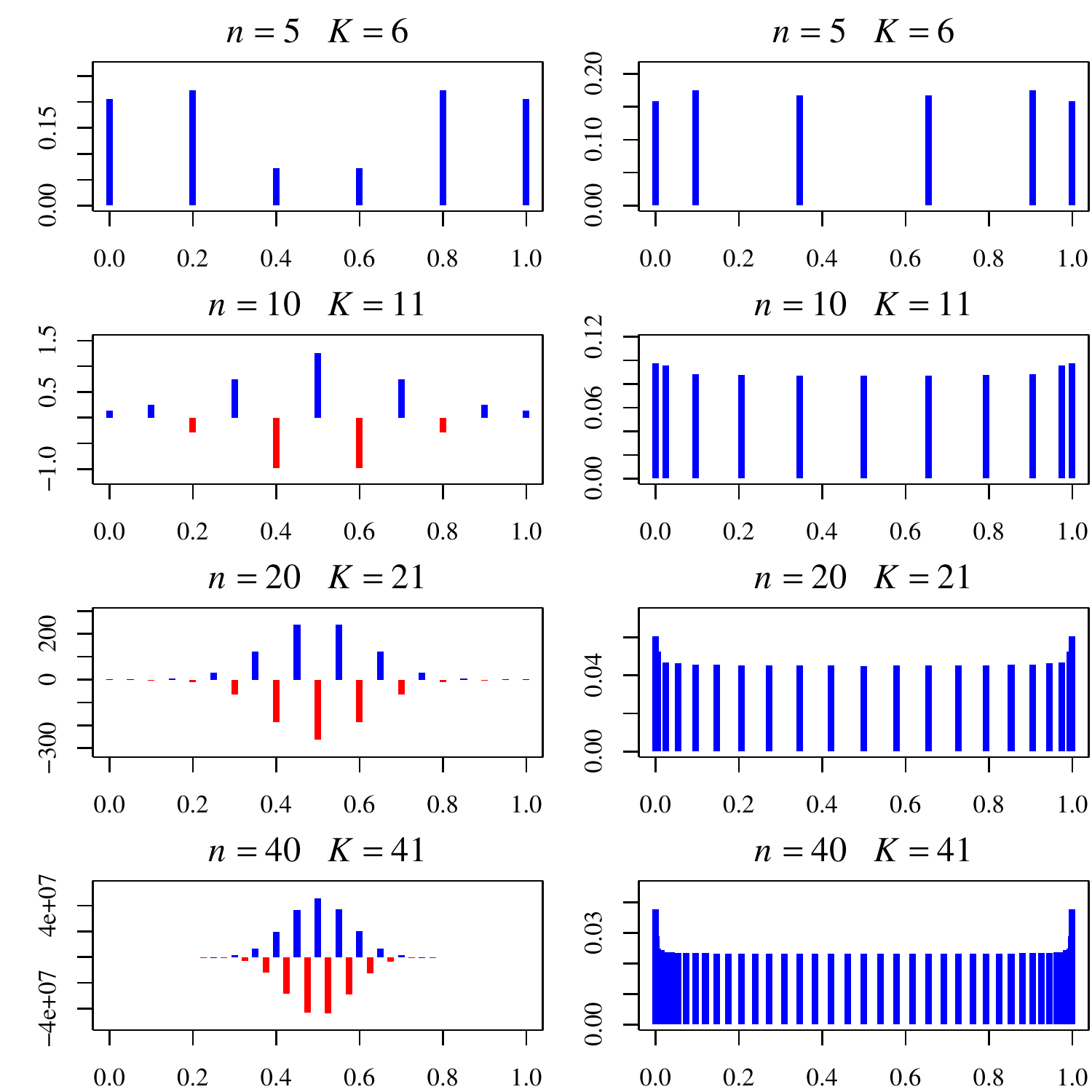}
\caption{Examples of priors representing the NML distribution in the
  Bernoulli model with $n=5,10,20,40$. The left panels show priors for
  $n+1$ mass points chosen at uniform intervals in $[0,1]$; the
  right panels show priors for the same number of mass points at
  $\sin^2((k-1)\pi/2n): k=1,\ldots,n+1$. The prior weights are
  obtained by directly solving a set of linear equations. Negative
  weights are plotted in red.}
\label{fig:linalg-plot}
\vspace{-5mm}
\end{figure}

Figure~\ref{fig:linalg-plot} shows priors representing NML obtained by
solving the associated linear equations. For mass points given by
$\sin^2((k-1)\pi/2n): k=1,\ldots,n+1$, the prior is nearly uniform
except at the boundaries of the parameter space where the weights are
higher.
For uniformly spaced mass points, the prior involves both
negative and positive weights when $n\geq 10$. Without the
non-negativity constraint, the requirement that the weights sum to one
no longer implies a bound ($W \leq 1$) on the magnitudes of the
weights, and in fact, the absolute values of the weights become
very large as $n$ grows.


\subsection{Bernoulli trials: Divergence optimization of $\theta$ and $W_n$}

For less than $n+1$ parameter values $\theta_k$ with non-zero prior
probability, there is no guarantee that an exact representation of NML
is possible. However, $\lfloor n/2 \rfloor +1$ mass points should suffice when we are also free to choose the $\theta$ values because the total degree of freedom of the symmetric discrete prior, $\lfloor n/2 \rfloor$, coincides with the number of equations to be satisfied by the mixture.

While solving the required weights, $W_{k,n}$, can be done in a
straightforward manner using linear algebra, the same doesn't hold for
$\theta_k$. Inspired by the work of Watanabe and Ikeda
{\cite{wi12}}, we implemented a Newton type algorithm
for minimizing
$$
   {1\over \beta} \log \sum_{x^n} \NML(x^n) \left({\NML(x^n) \over 
     \sum_{k=1}^{K} p(x^n\mid \theta_k) W_{k,n}}\right)^\beta
$$ with a large values of $\beta$, under the constraint that
$\sum_{k=1}^{K} W_{k,n} = 1$, where $K$ is the number of mass points. This optimization criterion is
equivalent to the Renyi divergence \cite{renyi}, and converges to
the log of the worst-case ratio
$$\log \max_{x^{n}}{\NML(x^n) \over \sum_{k=1}^K p(x^n\mid \theta_k)
W_{k,n}}$$ as $\beta\rightarrow \infty$.  In the following, we use
$\beta=150$ except for $n=500$ where we use $\beta=120$ in order to
avoid numerical problems. The mass points were initialized 
at $\sin^2((k-1)\pi/2(K-1)) : k=1,\ldots,K$.
   
\begin{figure}
\footnotesize
\hspace*{17mm}$K=\lfloor n/2 \rfloor + 1$\hspace*{27mm}$K=n+1$\\
  \includegraphics[width=\columnwidth]{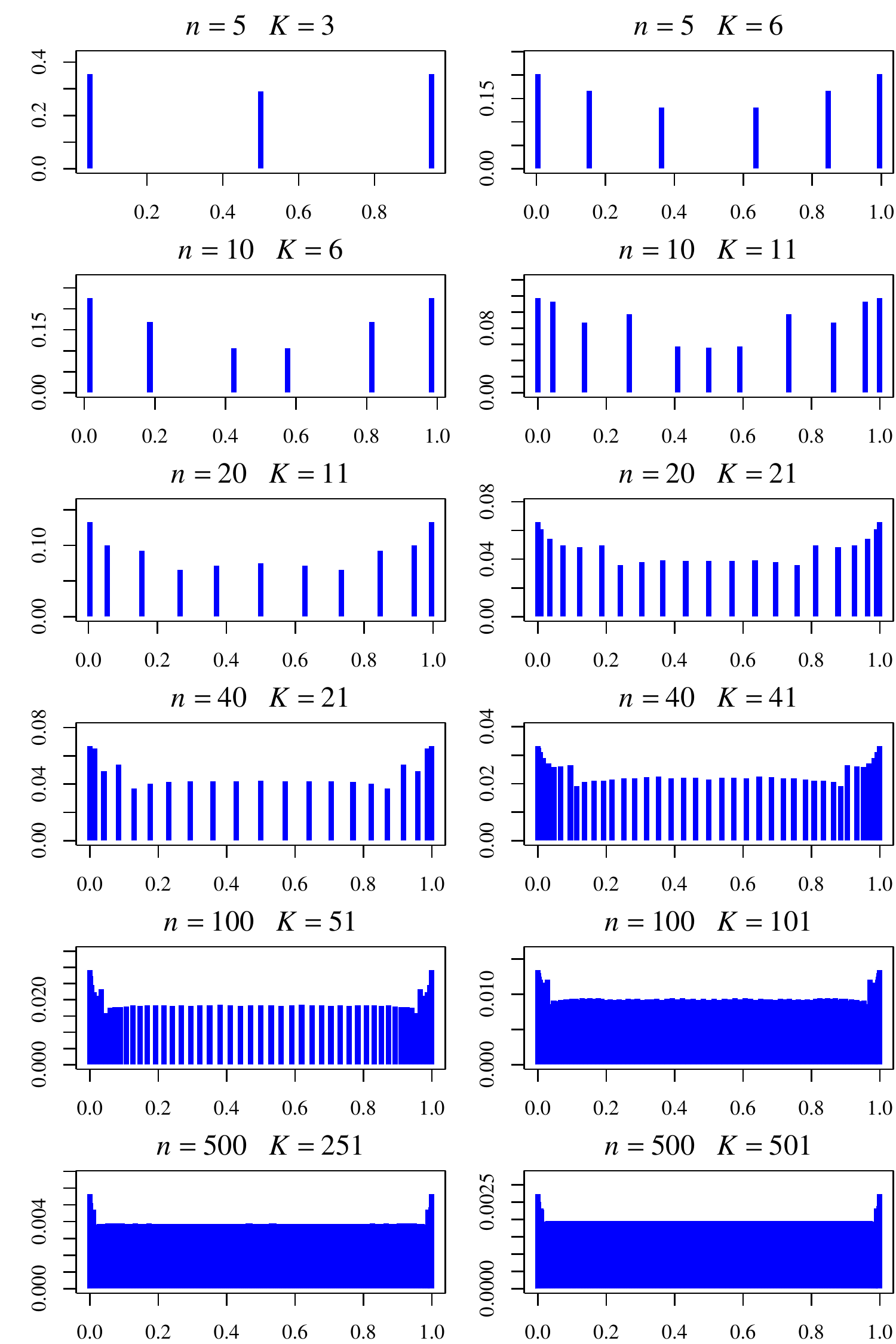}
\caption{Examples of priors representing the NML distribution in
  the Bernoulli model with $n=5,10,20,40,100,500$. 
  Both the locations of the mass points, $\theta_k$, and the 
  respective prior weights, $W_{k,n}$, are optimized using a Newton
  type method. The left panels show priors with $\lfloor n/2
  \rfloor +1$ mass points; the right panel shows priors with
  $n+1$ mass points.}
\label{fig:newton-plot}
\vspace{-5mm}
\end{figure}

Figure~\ref{fig:newton-plot} shows the priors obtained by optimizing
the locations of the mass points, $\theta_k$, and the respective prior
weights, $W_k$. The left panels show priors where the number of mass
points is $\lfloor n/2 \rfloor + 1$, while the
right panels shows priors with $n+1$ mass points which guarantees
that an exact representation is possible even without optimization of
the $\theta_k$.  Note, however, that the divergence optimization method we use can
only deal with non-negative prior weigths.
The obtained mixtures had Kullback-Leibler divergence $D(\NML||mix) <
10^{-7}$ and worst-case ratio $\max_{x^n} \NML(x^n)/mix(x^n) <
1+10^{-3}$ in each case.

\section{Conclusions and Future Work}

Unlike many earlier studies that have focused on either finite-sample
or asymptotic approximations of the normalized maximum likelihood
(NML) distribution, the focus of the present paper is on exact
representations.  We showed that an exact representation of NML as a
Bayes-like mixture with a possibly signed prior exists under a mild
condition related to linear independence of a subset of the
statistical model in consideration.
We presented two techniques for finding the required signed priors in
the case of Bernoulli trials.

The implications of this work are two-fold. First, from a theoretical
point of view, it provides insight into the relationship between
MDL and Bayesian methods by
demonstrating that in some models, a finite-sample Bayes-like
counterpart to NML only exists when the customary assumption that
prior probabilities are non-negative is removed. This complements
earlier asymptotic and approximate results. Second, from a practical
point of view, a Bayes-like representation offers a computationally
efficient way to extract marginal probabilities $p(x_1,\ldots,x_i)$
and conditional probabilities $p(x_i|x_1,\ldots,x_{i-1})$ for $i<n$
where $n$ is the total sample size. These probabilities are required
in, for instance, data data compression using arithmetic coding.  

Other algorithms will be explored in the full paper along with other
families including truncated Poisson and multinomial, which have an
interesting relationship between supports for the prior.  The full
paper will also show how matching NML produces a prior with some
interesting prediction properties. In particular, in Bernoulli trials,
by the NML matching device, we can arrange a prior in which the
posterior mean of the log odds is the same as the maximum likelihood
estimate of the log odds, whenever the count of ones is neither $0$
nor $n$.

\vspace{-2mm}

\ifCLASSOPTIONcaptionsoff
  \newpage
\fi


\begin{thebibliography}{99}



\bibitem{BarronRissanenYu}
A.\ R.\ Barron,
J.\ Rissanen
and
B.\ Yu,
``The minimum description length principle
in coding and modeling,''
{\it IEEE Trans.\ Inform.\ Theory,} 
Vol.\ 44 No.\ 6, pp.\ 2743 - 2760, 1998. 


\bibitem{bghhk} P.\ Bartlett,  P.\ Gr{\"{u}}nwald,  P.\ Harremo{\"{e}}s, F.\ Hedayati, W.\ Kot{\l}owski,
''Horizon-independent optimal prediction with log-loss in exponential families," arXiv:1305.4324v1, May 2013.  




\bibitem{ClarkeBarron} B.\ Clarke \& A.\ R.\ Barron, 
``Jeffreys prior is asymptotically least favorable under entropy risk,''
{\em J.\ Statistical Planning and Inference,} 41:37-60, 1994.

\bibitem{Grunwald} P.\ D.\ Gr{\"{u}}nwald, {\em The Minimum Description Length Principle},
MIT Press, 2007.

\bibitem{KontkanenMyllymaki} P.\ Kontkanen \& P.\ Myllym{\"{a}}ki.
``A linear-time algorithm for computing the multinomial stochastic complexity.''
{\it Information Processing Letters}, vol.103, pp.227-233, 2007.


\bibitem{Haussler} D.\ Haussler,
``A general minimax result for relative entropy,''
{\em IEEE Trans.\ Inform.\ Theory,}
vol.\ 43, no.\ 4, pp.\ 1276-1280, 1997.



\bibitem{renyi}
A.\ Renyi,
``On measures of entropy and information,''
{\it Proc.\ of the Fourth Berkeley Symp.\ on Math.\ Statist.\ and Prob.\,} 
vol. 1, Univ.\ of Calif.\ Press, pp.\  547-561, 1961.



\bibitem{Shtarkov} Yu M.\ Shtarkov,
``Universal sequential coding of single messages,''
{\em Problems of Information Transmission},
vol.\ 23, pp.\ 3-17, July 1988.

\bibitem{TakeuchiBarron98} J.\ Takeuchi \& A.\ R.\ Barron,
``Asymptotically minimax regret by Bayes mixtures,''
{\it Proc.\ 1998 IEEE ISIT,} 1998.

\bibitem{TakeuchiBarron13} J.\ Takeuchi \& A.\ R.\ Barron,
``Asymptotically minimax regret by Bayes mixtures for non-exponential families,'' 
{\it Proc. 2013 IEEE ITW}, pp.204-208, 2013.

\bibitem{wi12} K.\ Watanabe \& S.\ Ikeda,
``Convex formulation for nonparametric estimation of mixing distribution,''
{\it Proc.\ 2012 WITMSE,} pp.\ 36-39, 2012.

\bibitem{wrm} K.\ Watanabe, T.\ Roos \& P. Myllym\"aki,
``Achievability of asymptotic minimax regret in online and batch prediction,''
{\it Proc.\ 2013 ACML,} pp.\ 181-196, 2013.

\bibitem{XieBarron97}
Q.\ Xie \& A.\ R.\ Barron,
``Minimax redundancy for the class of memoryless sources'',
{\em IEEE Trans.\ Inform.\ Theory,} vol.\ 43, pp.\ 646-657, 1997.

\bibitem{XieBarron00}  Q.\ Xie \& A.\ R.\ Barron, 
``Asymptotic minimax regret for
data compression, gambling and prediction,'' 
{\em IEEE Trans.\ Inform.\ Theory,} vol.\ 46, pp.\ 431-445, 2000.




\end{thebibliography}
\end{document}